\documentclass[
11pt,
a4paper,
oneside, 
headinclude,footinclude, 
BCOR5mm,
]{article}
\usepackage[utf8]{inputenc}
\usepackage{amsmath}
\usepackage{subcaption}
\usepackage{caption}
\usepackage{longtable}
\usepackage{amsthm}
\usepackage{color}
\usepackage{sectsty}
\usepackage{hyperref}
\hypersetup{
    colorlinks=true, 
    linkcolor=blue, 
    urlcolor=red, 
    linktoc=all 
}
\usepackage{amssymb}
\usepackage{pagecolor,lipsum}
\usepackage{enumerate}
\usepackage{pdfpages}
\usepackage{xcolor}
\usepackage{tikzsymbols}
\usepackage{blkarray}
\usepackage{enumitem}
\usepackage{graphicx}
\usepackage{setspace}
\usepackage{authblk}
\usepackage{soul}
\usepackage{enumitem}
\usepackage{soul}
\usepackage{MnSymbol}
\usepackage{mathtools}
\usepackage{braket}
\usepackage{mdframed}
\usepackage{amsmath}
\usepackage[thinc]{esdiff}
\usepackage{array}
\usepackage{booktabs}
\DeclarePairedDelimiter\ceil{\lceil}{\rceil}

\usepackage[linesnumbered,ruled,vlined,english]{algorithm2e}
\usepackage[noend]{algpseudocode}
\usepackage[linesnumbered,ruled,vlined,english]{algorithm2e}
\usepackage{geometry}
\usepackage{tgtermes}

\let\oldpara\paragraph
\renewcommand{\paragraph}[1]{\vspace{-0.5cm}\oldpara{#1}}

\newcommand{\cnp}{\mbox{\textbf{NP}}}
\newcommand{\cp}{\mbox{\textbf{P}}}

\DeclareMathOperator*{\argmin}{arg\,min}

\newcommand{\mc}[1]{\mathcal{#1}}

\newcommand{\term}[1]{\textbf{\textit{#1}}}

\newcommand{\ts}[1]{\textsc{#1}}
\newcommand{\ep}[1]{\mathbb{E}}
\newcommand{\done}[1]{\textcolor{blue}{(\textbf{Done})}}

\newcommand{\pr}[1]{\mathbb{Pr}}

\newtheorem*{innerthm}{\innerthmname}
\newcommand{\innerthmname}{}
\newenvironment{statement}[1]
 {\renewcommand{\innerthmname}{#1}\innerthm}
 {\endinnerthm}
\theoremstyle{definition}

\newcommand{\z}[1]{\mathbb{Z}}
\newcommand{\nminfpe}[1]{\textsc{NMin-FPE}}
\newcommand{\nmaxfpe}[1]{\textsc{NMax-FPE}}
\newcommand{\nminfpr}[1]{\textsc{NMin-FPR}}
\newcommand{\nmaxfpr}[1]{\textsc{NMax-FPR}}

\newcommand{\zt}[1]{\mathbb{Z}_T}

\newcommand{\sydsG}{G_{\mc{S}}}

\newtheorem{theorem}{\textbf{Theorem}}[section]

\newtheorem{definition}[theorem]{\textbf{Definition}}

\newtheorem{proposition}[theorem]{Proposition}
\newtheorem{observation}[theorem]{Observation}

\definecolor{mycolor}{rgb}{0.122, 0.435, 0.698}
\newmdenv[topline=false, bottomline=false, rightline=false, innerlinewidth=0.4pt, roundcorner=4pt,linecolor=black,innerleftmargin=6pt,
innerrightmargin=6pt,innertopmargin=1pt,innerbottommargin=6pt]{mybox}

\newmdenv[backgroundcolor=gray!10, topline=false, bottomline=false, rightline=false, innerlinewidth=0.4pt, roundcorner=4pt,linecolor=black,innerleftmargin=6pt,
innerrightmargin=6pt,innertopmargin=3pt,innerbottommargin=6pt]{mybox2}

\newmdenv[backgroundcolor=blue!5, topline=false, bottomline=false, rightline=false, leftline=false, innerlinewidth=0.4pt, roundcorner=4pt,innerleftmargin=10pt,
innerrightmargin=10pt,innertopmargin=10pt,innerbottommargin=10pt]{mybox3}

\definecolor{darkblue}{RGB}{0,0,76}

\title{\textbf{Finding Nontrivial Minimum Fixed Points in \\ Discrete Dynamical Systems}}

\author { \small
    Zirou Qiu,\textsuperscript{1,2}
    Chen Chen,\textsuperscript{2}
    Madhav V. Marathe,\textsuperscript{1,2}
    S. S. Ravi,\textsuperscript{2,3}\\
    Daniel J. Rosenkrantz,\textsuperscript{2,3}
    Richard E. Stearns,\textsuperscript{2,3}
    Anil Vullikanti\textsuperscript{1,2}
}

\affil[1]{\small Computer Science Dept., University of Virginia.}
\affil[2]{\small Biocomplexity Institute and Initiative, University of Virginia.}
\affil[3]{\small Computer Science Dept., University at Albany – SUNY.}

\date{}

\begin{document}
\maketitle

\vspace{-1cm}
\begin{abstract}
\noindent
Networked discrete dynamical systems are often used to model the spread of contagions and decision-making by agents in coordination games. Fixed points of such dynamical systems represent configurations to which the system converges. In the dissemination of undesirable contagions (such as rumors and misinformation), convergence to fixed points with a small number of affected vertices is a desirable goal. Motivated by such considerations, we formulate a novel optimization problem of finding a nontrivial fixed point of the system with the minimum number of affected vertices. We establish that, unless \cp{} = \cnp{}, there is no polynomial time algorithm for approximating a solution to this problem to within the factor $n^{1 - \epsilon}$ for any constant $\epsilon > 0$. To cope with this computational intractability, we identify several special cases for which the problem can be solved efficiently. Further, we introduce an integer linear program to address the problem for networks of reasonable sizes. For solving the problem on larger networks, we propose a general heuristic framework along with greedy selection methods. Extensive experimental results on real-world networks demonstrate the effectiveness of the proposed heuristics.

\smallskip
\noindent
\textbf{Conference version.} The conference version of the paper is accepted at \texttt{\textbf{AAAI-2022}}: \href{https://ojs.aaai.org/index.php/AAAI/article/view/21174}{\textbf{Link}}.  
\end{abstract}

\section{Introduction}
Discrete dynamical systems are commonly used to model the propagation of contagions (e.g., rumors, failures of subsystems in infrastructures) and decision-making processes in networked games~\cite{valdez2020cascading, jackson2010social}. Specifically, the states of vertices in such dynamical systems are binary, with state~$1$ indicating the adoption of a contagion, and state $0$ otherwise. At each time step, the states of the vertices are updated using their \textit{local functions}. When the local functions are threshold functions, a vertex $v$ acquires a contagion (i.e., $v$ changes to state~$1$) if the number of $v$'s neighbors that have adopted the contagion (i.e., $v$'s peer strength) is at least a given threshold. Conversely, an individual's adoption of a contagion is {\em reversed} (i.e., $v$ changes to state~$0$) when the peer strength is below the threshold~\cite{barrett2006complexity}. Since its introduction by Granovetter~(\cite{granovetter1978threshold}), the threshold model has been extensively studied in many contexts including opinion dynamics~\cite{auletta2018reasoning}, information diffusion~\cite{cheng2018diffusion} and the spread of social conventions and rumors~\cite{dong2019multiple, ye2021collective}. The threshold model also captures decision patterns in networked coordination  games~\cite{ramazi2016networks}.


One important stage of the system dynamics is the convergence of vertices' states, where no individuals change states further; this is similar to an equilibrium in a networked game~\cite{daskalakis2007computing, daskalakis2015approximate}. Such a stage is called a \textbf{fixed point} of the dynamical system. Consider a scenario where a rumor is spreading in a community under the threshold model; here, an individual $v$ chooses to believe the rumor if the number of believers in $v$'s social circle is at least the threshold of $v$. Given the undesirable nature of rumors, identifying fixed points with minimum numbers of believers is desirable~\cite{wang2017drimux}.

\par  For some social contagions that are widely adopted in communities, it is often unrealistic to expect contagions to eventually disappear spontaneously. 
One example is the anti-vaccination opinion, which emerged in 1853 against the smallpox vaccine~\cite{wolfe2002anti}. Even today, the anti-vaccination sentiment persists across the world~\cite{willis2021covid}. 
Such considerations motivate us to study a more realistic problem, namely determining whether there are fixed points with at most a given number of contagion adoptions under the {\em nontriviality constraint} that the number of adoptions in the fixed point must be nonzero.
We refer to this as the \textbf{nontrivial minimum fixed point existence} problem~(\nminfpe{}).

Nontrivial minimum fixed points of a system, which are  jointly determined by the network structure and local functions, provide a way of quantifying the system's resilience against the spread of negative information. In particular, the number of contagion adoptions in a nontrivial minimum fixed point provides the lower bound on the number of  individuals affected by the negative contagion. 
Further, when the complete absence of a contagion is impractical, nontrivial minimum fixed points serve as desirable convergence points  for control strategies~\cite{khalil2013cuttingedge}. Similarly in coordination games, one is interested in finding equilibria wherein only a small number of players deviate from the strategy adopted by a majority of the players~\cite{ramazi2016networks}.


\par As we will show, the main difficulty of the \nminfpe{} problem lies in its computational complexity. 
A related problem is that of influence 
minimization (e.g., \cite{yao2015topic}). The main differences between the two problems are twofold. First, the influence minimization problem is based on the progressive model where a vertex state can only change from 0 to 1 but not vice versa. Second, the influence minimization problem aims to find optimal intervention strategies (e.g., vertex/edge removal) to reduce the cascade size, while \nminfpe{} aims to find a minimum influenced group without changing the system. In this work, we study the \nminfpe{} problem on synchronous dynamical systems (SyDS) with threshold local functions, where the vertices update states simultaneously in each time-step. Our main contributions are as follows:
\begin{enumerate}[leftmargin=*,noitemsep,topsep=0pt]
    \item \textbf{Formulation.} We formally define the Nontrivial Minimum Fixed Point Existence Problem (\nminfpe{}) from a combinatorial optimization perspective. 
    
    \item \textbf{Intractability.} We establish that unless \textbf{P} $=$ \textbf{NP}, \nminfpe{} cannot be approximated to within the factor $n^{1 - \epsilon}$ for any $\epsilon > 0$, even when the graph is bipartite. We also show that the \nminfpe{} is \textbf{W[1]}-hard w.r.t. the natural parameter of the problem
    (i.e., the number of vertices in state 1 in any nontrivial fixed point).
    
    \item \textbf{Algorithms.} We identify several special cases for which \nminfpe{} can be solved in polynomial time. 
    To obtain an optimal solutions for networks of moderate size, we present an integer linear program (ILP) formulation for \nminfpe{}. For larger networks, we  propose a heuristic framework along with three greedy selection strategies that can be embedded into the framework.
    
    \item \textbf{Evaluation.} We conduct extensive experiments to study the performance of our heuristics on real-world and synthetic networks under various scenarios. Our results demonstrate that the proposed heuristics are effective, and the proposed method outperforms baseline methods significantly, despite the strong inapproximability of \nminfpe{}.
\end{enumerate}

\section{Related work}
\medskip
\paragraph{Fixed points.} Fixed points of discrete dynamical systems have been widely studied. Goles and Martinez \cite{goles2013neural} show that for any initial configuration, a threshold SyDS always converges to either
a fixed point or a cycle with two configurations in a polynomial number of time steps. Barrett et al.~\cite{BH+07} show that determining whether a system has a fixed point (\ts{FPE}) is \cnp-complete for symmetric sequential dynamical systems and that the problem is efficiently solvable for threshold sequential dynamical systems. 
More recently, Chistikov et al. \cite{chistikov2020convergence} study fixed points in the context of opinion diffusion;  they show that determining whether a system reaches a fixed point from a given configuration is \textbf{PSPACE}-complete for SyDSs on general directed networks, but can be solved in polynomial time when the underlying graph is a DAG. In Rosenkrantz et al. \cite{rosenkrantz2020synchronous}, we investigate convergence and other problems for SyDSs whose underlying graphs are DAGs.  In particular, we show that the 
convergence guarantee problem (i.e., determining if a system reaches a fixed point starting from any configuration) is
\textbf{Co}-\textbf{NP}-complete for SyDSs on DAGs.

\paragraph{Influence minimization.} Existing works on influence minimization focus on reducing the prevalence via control strategies. Yang, Li and Giua~\cite{yang2019influence} study the problem of finding a $k$-subset of active vertices
(i.e., an initial configuration with at most $k$ vertices in state 1) such that the converged influence value is minimized and a target set of vertices are active. They provide an integer program of the problem and suggest two heuristics. Wang et al.~\cite{wang2017drimux}  propose a new rumor diffusion model and optimize blocking the contagion by considering an Ising model. Zhu, Ni, and Wang~\cite{zhu2020activity} estimate the influence of vertices and minimize the adoption of negative contagions by disabling vertices.  Other approaches focus on blocking the spread via vertex removal~\cite{kimura2007extracting, yao2015topic,chen2015vertex,kuhlan-etal-2015} or edge removal~\cite{kimura2008minimizing, khalil2013cuttingedge, chen2016eigen,kuhlman-etal-2013} 
and enhance network resilience~\cite{ chen2015vertex}.

\paragraph{Coordination games.} Agent decision-making in coordination games coincides with threshold-based cascade of contagions. Adam et al.~\cite{adam2012behavior} study the best response dynamics of coordination games and analyze the convergences and propose a new network resilience measure. 
Ramazi et al. (\cite{ramazi2016networks}) study both coordination and anticoordination games and show that such games always reach equilibria in a finite amount of time. Other aspects of equilibria in networked games 
(such as developing control strategies and determining the existence of equilibria) have also been
studied~\cite{yu2020computing, cao2008minimax, anderson2001minimum, salehisadaghiani2018distributed}.

\section{Preliminaries and Problem Definition}

We follow the definition of discrete dynamical systems from previous work~\cite{rosenkrantz2020synchronous}. A \textbf{synchronous dynamical system} (SyDS) $\mc{S}$ over the Boolean domain $\mathbb{B} = \{0, 1\}$ of state values is defined as a pair $(\sydsG{}, \mc{F})$ where ($1$) $\sydsG{} = (V, E)$ is the underlying graph of $\mc{S}$ with $n = |V|$ and $m = |E|$, and ($2$) $\mc{F} = \{f_1, ..., f_n\}$ is a collection of functions for which $f_i$ is the \textit{local transition function} of vertex $v_i \in V, 1 \leq i \leq n$. In general, $f_i \in \mc{F}$ specifies how $v_i \in V$ updates its state throughout the evolution of $\mc{S}$. In this work, we study SyDSs over the Boolean domain with threshold functions as local functions.
Following~\cite{barrett2006complexity},
We denote such a system by $(\ts{Bool}, \ts{Thresh})$-SyDS.

\paragraph{Update rules.} In $(\ts{Bool}, \ts{Thresh})$-SyDSs, each vertex $v_i \in V$ has a fixed integer threshold value $\tau_{v_i} \geq 0$. At each time step $t \geq 0$, each vertex $v_i \in V$ has a state value in $\mathbb{B}$. While the initial state of any vertex $v_i$ (at time $0$) can be assigned arbitrarily, the states at time steps $t \geq 1$ are determined by $v_i$'s local function $f_i$. Specifically, $v_i$ transitions to state $1$ at time $t$ if the number of state-1 vertices in its closed neighborhood $N(v_i)$ (which consists of $v_i$ and all its neighbors) at time $t-1$ is at least $\tau_{v_i}$; the state of $v_i$ at time $t$ is $0$ otherwise. Furthermore, all vertices update their states synchronously. When $\sydsG{}$ is a directed graph, a vertex $v$ transitions to state $1$ at a time $t \geq 1$ iff the number of state-$1$ \textit{in-neighbors} of $v$
(i.e., vertex $v$ itself and those from which $v$ has
incoming edges) 
is at least $\tau_v$. Note that undirected networks are the primary focuses of this work; we assume that $\sydsG{}$ is undirected unless specified otherwise.

\begin{figure}[!h]
\small
  \centering
    \includegraphics[width=0.7\textwidth]{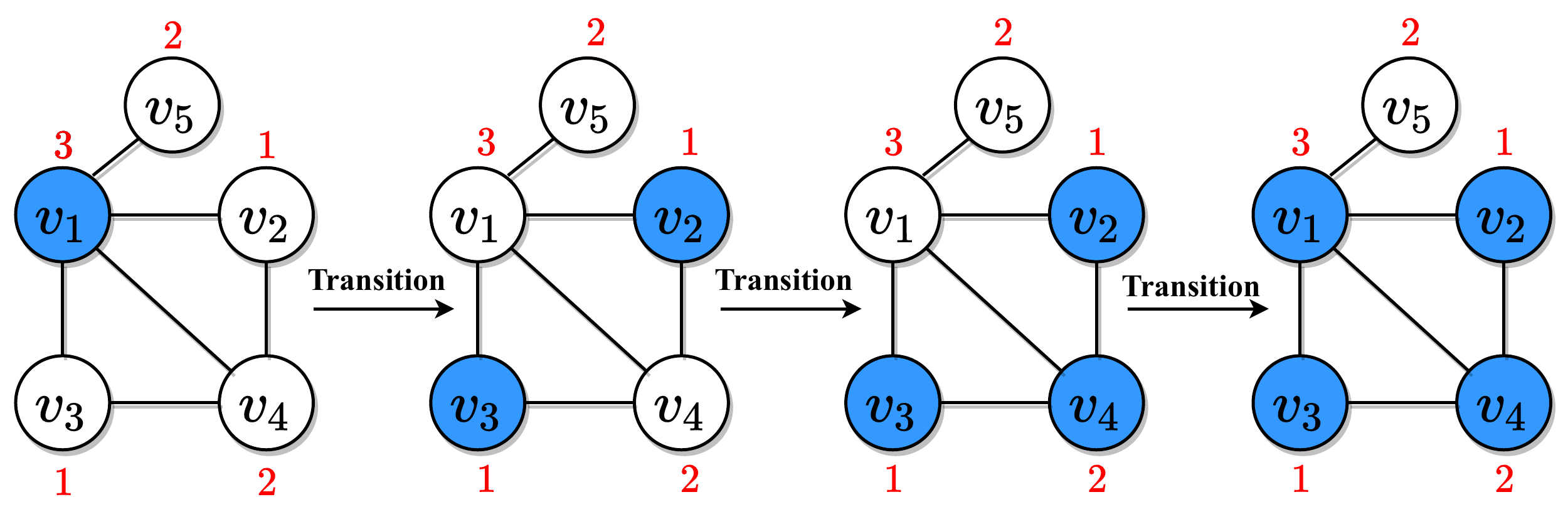}
    
    \caption{The evolution of a $(\ts{Bool}, \ts{Thresh})$-SyDS $\mc{S} = (\sydsG{}, \mc{F})$ where $V(\sydsG{}) = \{v_i : i = 1, ..., 5\}$. The threshold values are shown in red: $\tau_1 = 3, \tau_2 = 1, \tau_3 = 1, \tau_4 = 2,$ and $\tau_5 = 2$. State-1 vertices are highlighted in blue. The system undergoes the following evolution: $(1, 0, 0, 0, 0) \rightarrow (0, 1, 1, 0, 0) \rightarrow (0, 1, 1, 1, 0) \rightarrow (1, 1, 1, 1, 0)$, with the last configuration being a fixed point.}
    
    \label{fig:example_syds}
\end{figure}

\paragraph{Configurations and fixed points.} 
A \textbf{configuration} of $\mc{S}$ gives the states of all vertices during a time-step. Specifically, a configuration $C$ is an $n$-vector $C =$ $(C(v_1), C(v_2), ..., C(v_n))$ where $C(v_i) \in \mathbb{B}$ is the state of vertex $v_i \in V$ under $C$. There are a total of $2^n$ possible configurations a system $\mc{S}$. During the evolution of the $\mc{S}$, the system configuration $C$ changes over time. If $\mc{S}$ transitions from $C$ to $C'$ in one time step, then $C'$ is the \textit{successor} of $C$. Due to the deterministic nature of $(\ts{Bool}, \ts{Thresh})$-SyDSs, if $C = C'$, that is, the states of all vertices remain unchanged, then $C$ is a \textbf{fixed point} of the system. An example of a $(\ts{Bool}, \ts{Thresh})$-SyDS $\mc{S} = (\sydsG{}, \mc{F})$ is shown in Figure~\ref{fig:example_syds}. Note that given any configuration $C$, its successor $C'$ 
can be computed in time that is polynomial in the number of vertices $n$.  As shown in~\cite{goles2013neural}, starting from any initial configuration, $\mc{S}$ converges either to a fixed point or a cycle consisting of two configurations within a number of transitions that is a polynomial in $n$. 

\paragraph{Constant-state vertices.} A vertex $v \in V$ is a \textit{constant-1} vertex if $\tau_v = 0$, that is, given any configuration $C$,  the state of $v$ is 1 in the successor $C'$ of $C$. Similarly, $v$ is a \textit{constant-0} vertex if $\tau_v = deg(v) + 2$.

\subsection{Problem definition}

\par Let $\mc{S} = (\sydsG{}, \mc{F})$ be a SyDS and let $C$ be a configuration of $\mc{S}$. The \textit{Hamming weight} of $C$, denoted by $H(C)$, is the number of $1$'s in $C$. A \textit{minimum fixed point} of $\mc{S}$ is a fixed point with the smallest possible Hamming weight. 
Note that when a $(\ts{Bool}, \ts{Thresh})$-SyDS $\mc{S}$ has no constant-1 vertices, the minimum fixed point of $\mc{S}$ is trivially $\textbf{0}^n$. 
A \textit{nontrivial fixed point} is a fixed point that is different from $\textbf{0}^n$.
Our work focuses on finding \textit{nontrivial minimum fixed points}.

\begin{mybox2}
\begin{definition}
A \textbf{nontrivial minimum fixed point} of $\mc{S}$ is a nontrivial fixed point of minimum Hamming weight.
\end{definition}
\end{mybox2}

\vspace{-5px}
\noindent
We now provide a formal definition of the problem.

\begin{mybox2}
\textbf{Nontrivial Minimum Fixed Point Existence} (\nminfpe{})

\noindent
\underline{Instance:} A SyDS $\mc{S} = (\sydsG{}, \mc{F})$ and a positive integer $q$.

\noindent
\underline{Question:} Is there a fixed point $C$ of $\mc{S}$ with Hamming weight at least $1$ and at most $q$?
\end{mybox2}

\noindent
We focus on the \textbf{NP}-optimization version of \nminfpe{} which is to find a nontrivial minimum fixed point.

\section{Computational Hardness of NMIN-FPE}
\label{sec:nontrivial-min-1-fpe}
In this section, we present an inapproximability result for \nminfpe{}. Specifically, we show that \nminfpe{} cannot be poly-time approximated within a factor $n^{1 - \epsilon}$ for any constant $\epsilon > 0$, unless $\textbf{P} = \textbf{NP}$. We also establish that \nminfpe{} is \textbf{W[1]}-hard, with the parameter being the Hamming weight of a fixed point. 
Under standard hypotheses in computational complexity, our
results rule out the possibility of obtaining efficient approximation algorithms with provable performance guarantees
and fixed parameter tractable algorithms w.r.t. the Hamming weight for \nminfpe{}.

\begin{mybox2}
\begin{theorem}\label{thm:hard-to-approx}
The problem \nminfpe{} \textbf{cannot} be approximated to within a factor $n^{1 - \epsilon}$ for any constant $\epsilon > 0$, unless $\textbf{P} = \textbf{NP}$. This inapproximability holds even when the underlying graph is bipartite.
\end{theorem}
\end{mybox2}

\begin{figure}[!h] 
  \centering
    \includegraphics[width=0.7\textwidth]{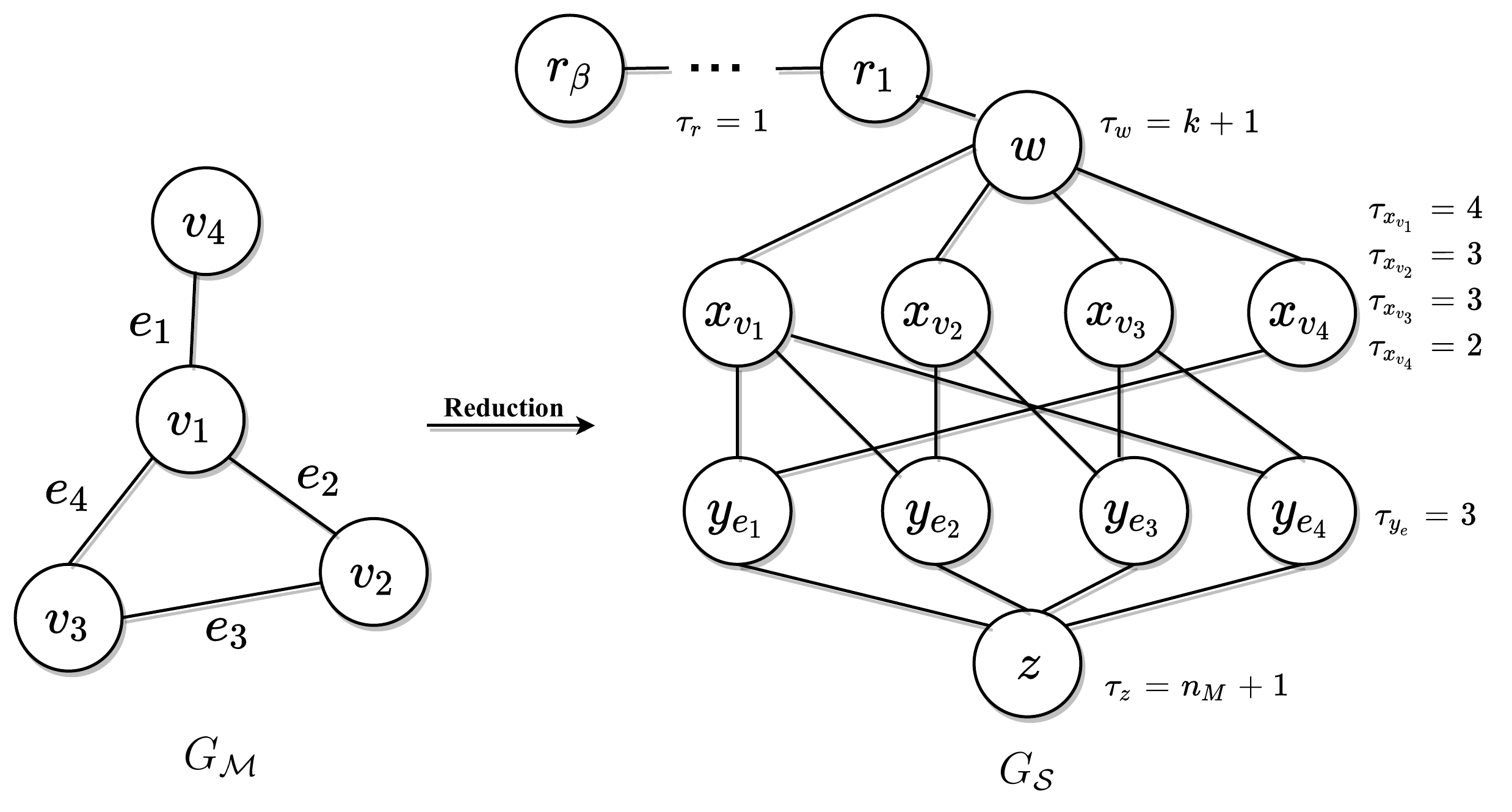}
    \caption{An example of the reduction from \ts{MVC} to \nminfpe{} where $G_\mc{M}$ and $G_{\mc{S}}$ are graphs for \ts{MVC} and \nminfpe{}, respectively.}
    \label{fig:example_reduction_sup}
\end{figure}

\begin{proof}
We first consider the case where $0 < \epsilon < 1$. The overall scheme of our proof is a reduction from \ts{minimum vertex cover} to \nminfpe{} such that if there exists a poly-time factor $n^{1- \epsilon}$ approximation algorithm $\mc{A}$ for \nminfpe{}, we then can use $\mc{A}$ to solve the \ts{Minimum vertex cover} problem in polynomial time, implying $\textbf{P} = \textbf{NP}$. 
Let $\mc{M} = \langle G_{\mc{M}}, k \rangle$ be an arbitrary instance  of \ts{Minimum vertex cover} (\ts{MVC}) with target size $k$, where $n_{\mc{M}} = |V(G_\mc{M})|$ and $m_{\mc{M}} = |E(G_\mc{M})|$. Without loss of generality, we assume that $G_{\mc{M}}$ is connected. Moreover, observe when $k = n_{\mc{M}} - 1$, \ts{MVC} can be solved in polynomial time. Therefore, we assume that $k < n_{\mc{M}}-1$. 

\par Let $\mc{A}$ be a \textbf{polynomial} time factor $n^{1 - \epsilon}$ approximation algorithm for \nminfpe{}, $0 < \epsilon < 1$, where $n$ is the number of vertices in the underlying graph. We build an instance $\mc{S} = (\sydsG{}, \mc{F})$ of \nminfpe{} for which $n_{\mc{S}} = |V(\sydsG{})|$ and $m_{\mc{S}} = |E(\sydsG{})|$. Let $\alpha = m_{\mc{M}} + n_{\mc{M}} + 1$, and $\beta = \alpha^{\ceil{2/\epsilon}}$. The construction is as follows.

\begin{enumerate}
[leftmargin=*,noitemsep,topsep=0pt]
    \item \textbf{The vertex set $V(G_{\mc{S}})$.} Let $X = \{x_u : u \in V(G_\mc{M})\}$ and  $Y = \{y_e : e \in E(G_\mc{M})\}$ be two sets of vertices that corresponds to vertices and edges in $G_{\mc{M}}$, respectively. Let $w, z$ be two additional vertices. Lastly, we introduce a set of $\beta$ vertices $R = \{r_1, ..., r_{\beta}\}$. Then $V(G_{\mc{S}}) = X \cup Y \cup R \cup \{w\} \cup \{z\}$.
    
    \par We use notations $x_u \in X$, $y_e \in Y$, $r_i \in R$, $w$ and $z$ and to distinguish the five classes of vertices in $G_{\mc{S}}$.
    
    \item \textbf{The edge set $E(G_{\mc{S}})$.} Let $E_1 = \{(x_u, y_e) :$ $u \in V(G_{\mc{M}})$ is incident to $e \in E(G_{\mc{M}}) \}$ be the edge set such that if a vertex $u$ and an edge $e$ are incident in $G_{\mc{M}}$, their corresponding vertices $x_u$ and $y_e$ are adjacent in $G_{\mc{S}}$. Let $E_2 = \{ (y_e, z) : y_e \in Y\}$ be the edge set where vertex $z$ is adjacent to all $y_e \in Y$. Let $E_3 = \{(x_u, w) : x_u \in X\}$ be the edge set such that $w$ is adjacent to all $x_u \in X$. Let $E_4 = \{(r_i, r_{i+1}) : r_i \in R, 1 \leq i \leq \beta - 1\}$, that is, vertices in $R$ form a path. Lastly, we introduce an additional edge $(w, r_1)$ to connect $w$ with an endpoint of the above path. The edge set of $G_{\mc{S}}$ is then defined as $E(\sydsG{}) = \bigcup_{i=1}^4 E_i \cup \{(w, r_1)\}$. 
    
    \item \textbf{Thresholds.} The threshold $\tau_{x_u}$ of each $x_u \in X$ is $deg(u) + 1$ where $deg(u)$ is the degree of vertex $u$ in $G_{\mc{M}}$. The thresholds $\tau_{y_e}$ of all $y_e \in Y$ is $3$. The thresholds of $w$ and $z$ are $k + 1$ and $m_{\mc{M}} + 1$, respectively. Lastly, all $r_i \in R$ have threshold $1$. 
\end{enumerate}
\noindent
This completes the construction of the \nminfpe{} instance $\mc{S} = (G_{\mc{S}}, \mc{F})$ which clearly takes polynomial time w.r.t. $n_{\mc{M}}$. The resulting graph $\sydsG{}$ has $n_{\mc{S}} = n_{\mc{M}} + m_{\mc{M}} + \beta + 2$ vertices and $m_{\mc{S}} = n_{\mc{M}} + 3 m_{\mc{M}} + \beta$ edges. Furthermore, $\sydsG{}$ is bipartite. An example of reduction is shown in Figure~\ref{fig:example_reduction_sup}. We now argue that $G_{\mc{M}}$ has a vertex cover of size at most $k$ if and only if the algorithm $\mc{A}$ returns a non-zero fixed point of size at most $n_{\mc{S}}^{1 - \epsilon} \alpha$ where $\alpha = m_{\mc{M}} + n_{\mc{M}} + 1$. 

\begin{itemize}
    \item[($\Rightarrow$)] 
    Let $V' \subseteq V(G_{\mc{M}})$ be a vertex cover of size $k$. We can construct a fixed point $I$ of $\mc{S}$ by setting (1) the states of all $y_e \in Y$ to $1$, (2) for all $u \in V'$, the states of $x_u \in X$ to $1$, and (3) the state of vertex $z$ to 1. The number of 1's in the resulting configuration $I$ is $m_{\mc{M}} + k + 1$. We now show that $I$ is a fixed point. Let $I'$ be the successor of $I$. It is easy to see that the state of vertices in $R \cup \{w\}$ remains $0$ and state of $z$ remains 1 in $I'$. We consider the following two cases of the state of a vertex $x_u \in X$ under $I$. If the state of $x_u$ is 1, then since all $y_e$'s have state $1$, the input to $x_u$'s local function is $deg(x_u) + 1 = \tau_{x_u}$. Thus, $x_u$ remains in state $1$ under $I'$. If the state of $x_u$ is $0$ under $I$, the input to $x_u$'s local function is $deg(x_u) < \tau_{x_u}$ and $x_u$ remains in state $0$ under $I'$. It follows that $I$ is a fixed point.
    
    Let $\ts{OPT}(\mc{S})$ denote the Hamming weight of a nontrivial minimum fixed point of $\mc{S}$ which is at most $k + m_{\mc{M}} + 1$ Let $C$ be a fixed point returned by algorithm $\mc{A}$. Since $\mc{A}$ have an approximation factor of $n^{1 - \epsilon}$, we have
    \begin{equation}
      H(C) \leq n_{\mc{S}}^{1 - \epsilon} \cdot \ts{OPT} \leq n_{\mc{S}}^{1 - \epsilon} \cdot (k + m_{\mc{M}} + 1) \leq n_{\mc{S}}^{1 - \epsilon} \cdot \alpha  
    \end{equation}
    
    \item[($\Leftarrow$)] We prove the contrapositive. Suppose $G_{\mc{M}}$ does not have a vertex cover of size at most $k$, we establish the following two claims.
    
    \begin{statement}{Claim 3.1.1} \label{claim:y_must_one}
    Under any fixed point $C$ of $\mc{S}$ where the vertex $w$ is in state $0$, there exists a state-1 vertex in $X \cup Y \cup \{z\}$ if and only if all $y_e \in Y$'s have state $1$ under $C$.
    \end{statement}
    The necessity of the claim clearly holds. We prove the contrapositive of the other direction. Suppose there exists a vertex $y_e \in Y$ with state $0$ under $C$, let $x_u$ and $x_v$ be the two neighbors of $y_e$. Observe $x_u$ and $x_v$ must have state $0$ since their thresholds are degrees plus one. Similarly, the vertex $z$ also has state $0$ under $C$. As a result, all neighbors of $x_u$ and $x_v$ have state $0$. By recursion, all vertices in $X \cup Y \cup \{z\}$ have state $0$ under $C$. This concludes the claim.
    
    \begin{statement}{Claim 3.1.2}
    For any non-zero fixed point $C$ of $\mc{S}$, all vertices in $R$ must have state $1$ under $C$.
    \end{statement}
    
    Suppose there exists a non-zero fixed point $C$ where at least one vertex $r_i \in R$ has state $0$. Since $\tau_{r_i} = 1, \; \forall r_i \in R$, it follows that all vertices in $R$ and vertex $w$ must have state $0$ under $C$. We now consider the states of other vertices under $C$. Given the $\tau_w = k + 1$, the number of state-$1$ vertices in $X$ is at most $k$. Note that however, since $G_{\mc{M}}$ does not have a vertex cover of size $k$, under any combinations of $k$ state-$1$ vertices in $X$, there exists at least one vertex $y_e \in Y$ such that both $y_e$'s neighbors in $X$ have state $0$ under $C$. Since $\tau_{y_e} = 3, \; \forall y_e \in Y$, there must exist a vertex $y_e \in Y$ with state $0$ under $C$. Follows from the Claim 3.1.1, $C$ is a zero fixed point which yields the contraposition. This concludes the proof of the claim.
    
    \par By the claim above, we have $\ts{OPT}(\mc{S}) \geq \beta = \alpha^{\ceil{2/ \epsilon}}$. We now argue that  $n_{\mc{S}}^{1 - \epsilon} \cdot \alpha < \alpha^{\ceil{2/ \epsilon}}$. Let $\lambda = \ceil{2 / \epsilon} - 2 / \epsilon$, we then rewrite $n_{\mc{S}}^{1 - \epsilon} \cdot \alpha$ as 
    \begin{equation}
        n_{\mc{S}}^{1 - \epsilon} \cdot \alpha = (\alpha^{\frac{2}{\epsilon} + \lambda} + \alpha + 1)^{1 - \epsilon} \cdot \alpha
    \end{equation}
    
    where $n_{\mc{S}} =  n_{\mc{M}} + m_{\mc{M}} + \beta + 2 = \alpha + \alpha^{\ceil{2 / \epsilon}} + 1$. Since $n_{\mc{M}} \geq 1$ and $0 \leq \epsilon < 1$, we have $\alpha \geq 2$ and $2 / \epsilon + \lambda = \ceil{2 / \epsilon} \geq 3$. Thus, 
    
    \begin{equation}
        \alpha + 1 < \alpha^{\frac{2}{\epsilon} + \lambda} \leq (\alpha - 1) \cdot \alpha^{\frac{2}{\epsilon} + \lambda}
    \end{equation}
    
    and it follows that
    \begin{equation}
     \alpha^{\frac{2}{\epsilon} + \lambda} + \alpha + 1 < \alpha^{\frac{2}{\epsilon} + \lambda + 1}
    \end{equation}
    
    Since $1 - \epsilon > 0$, 
    \begin{align*}
        (\alpha^{\frac{2}{\epsilon} + \lambda} + \alpha + 1)^{1 - \epsilon} \cdot \alpha &< (\alpha^{\frac{2}{\epsilon} + \lambda + 1})^{1 - \epsilon} \alpha \\
        &= \alpha^{(\frac{2}{\epsilon} + \lambda + 1)(1 - \epsilon) + 1}\\
        &= \alpha^{\ceil{\frac{2}{\epsilon}} - \epsilon \lambda - \epsilon}\\
        &< \alpha^{\ceil{\frac{2}{\epsilon}}}
    \end{align*}

\par Overall, we have $n_{\mc{S}}^{1 - \epsilon} \cdot \alpha < \alpha^{\ceil{2/ \epsilon}} \leq \ts{OPT}(\mc{S})$. Let $C$ be the non-zero fixed point returned by $\mc{A}$. It follows immediately that $H(C) > n_{\mc{S}}^{1 - \epsilon} \cdot \alpha$.
\end{itemize}

\noindent
Since $n_{\mc{S}} = n_{\mc{M}} + m_{\mc{M}} + \beta + 1$ is polynomial w.r.t. $n_{\mc{M}}$, the algorithm $\mc{A}$ runs in polynomial time w.r.t. $n_{\mc{M}}$. The reduction implies that if there exists such an algorithm $\mc{A}$ with approximation factor $n^{1 - \epsilon}$ for any constant $0 <\epsilon < 1$, based on the Hamming weight of the fixed point returned by $\mc{A}$, we can solve \ts{MVC} in polynomial time, implying $\textbf{P} = \textbf{NP}$. For the case where $\epsilon \geq 1$, note that the resulting approximation factor $n^{1-\epsilon}$ is less than any factor when $0 < \epsilon < 1$, thus, the inapproximability follows immediately. Overall, we conclude that \nminfpe{} cannot be approximated within a factor $n^{1 - \epsilon}$ for any constant $\epsilon > 0$, unless $\textbf{P} = \textbf{NP}$, and the hardness holds even when the underlying graph is bipartite.
\end{proof}

\par Theorem~(\ref{thm:hard-to-approx}) establishes a strong inapproximability result for \nminfpe{}; it points out that even obtaining an approximation guarantee that is slightly better than a linear factor is hard.

\paragraph{Parameterized Complexity of NMIN-FPE.}
We now examine whether \nminfpe{} is fixed-parameter tractable (FPT) w.r.t. a \textit{natural} structural parameter of the problem, namely the Hamming weight of a fixed point.

\begin{mybox2}
\begin{theorem}\label{thm:w1}
The problem \nminfpe{} is \textbf{W[1]-hard} w.r.t. to the natural parameter (i.e., the Hamming weight of a
fixed point) for (\ts{Bool, Thresh})-SyDSs.
\end{theorem}
\end{mybox2}

\begin{figure}[!h] 
  \centering
    \includegraphics[width=0.65\textwidth]{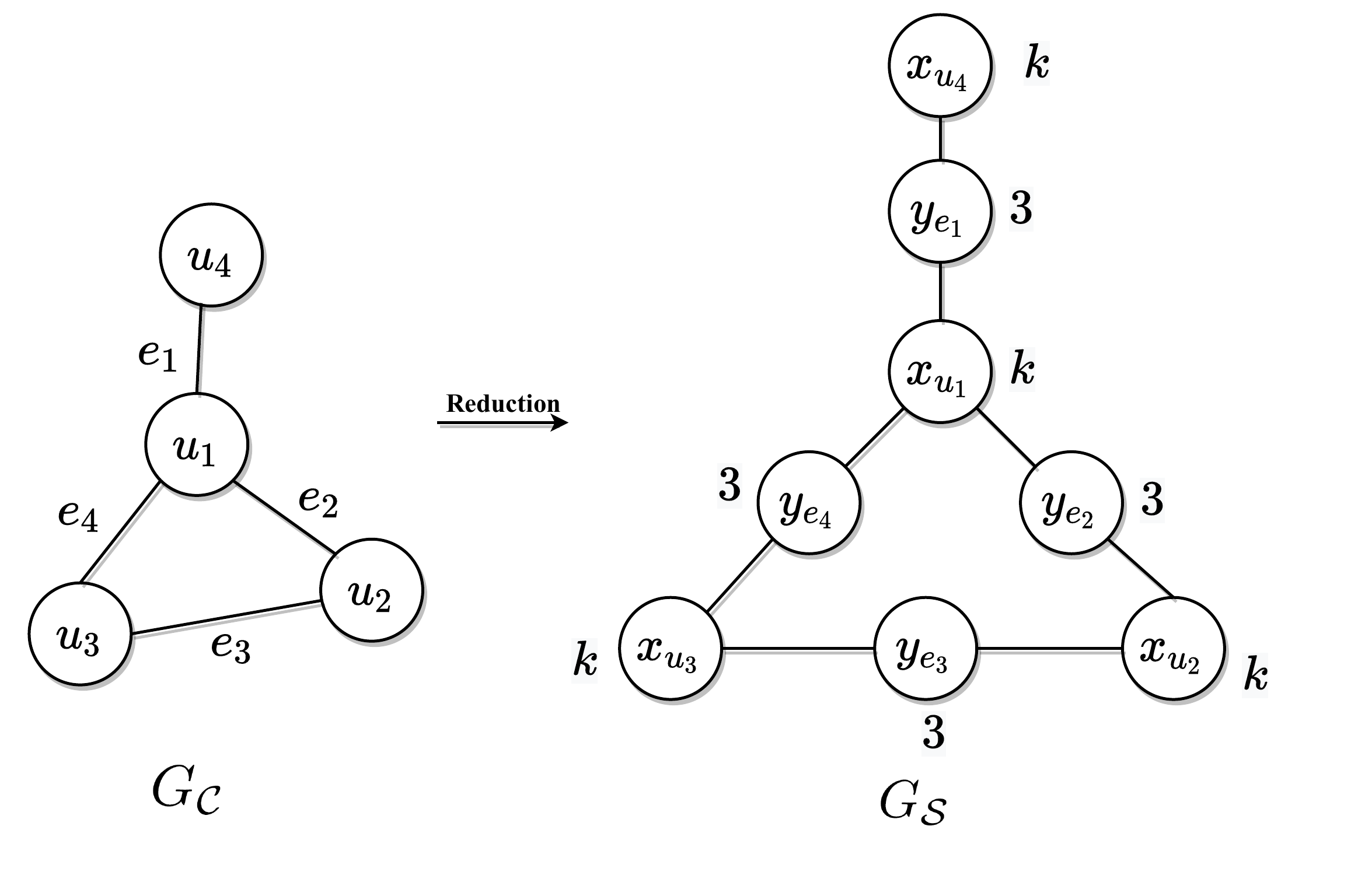}
    \caption{An example of the W[1]-hardness reduction from \ts{Clique} to \nminfpe{} where $G_\mc{C}$ and $G_{\mc{S}}$ are graphs for \ts{Clique} and \nminfpe{}, respectively. Threshold values are placed next to vertices in $G_{\mc{S}}$}
    \label{fig:w1-hard}
\end{figure}

\begin{proof}
We give a \textit{parameterized reduction} from the \ts{Clique} problem to \nminfpe{}. Let $\langle G_{\mc{C}}, k \rangle$ be an arbitrary instance of the \ts{clique} problem, we construct a \nminfpe{} instance $\langle \mc{S} = (G_{\mc{S}}, \mc{F}), q \rangle$ by generating a subdivision of $G_{\mc{C}}$. In particular, each edge $e = (u, v) \in E(G_{\mc{C}})$ is transformed into two edges $(x_u, y_e)$ and $(y_e, x_v)$ in $E(G_{\mc{S}})$, where (1) $x_u, x_v \in V(G_{\mc{S}})$ correspond to the vertices $u, v \in V(G_{\mc{C}})$, respectively, and (2) $y_e \in V(G_{\mc{S}})$ corresponds to the edge $e \in E(G_{\mc{C}})$. For clarity, we use $X = \{x_u : u \in V(G_{\mc{C}})\}$ and $Y = \{y_e : e \in E(G_{\mc{C}})\}$ to denote the sets of two types of vertices in $G_{\mc{S}}$ that correspond to vertices and edges in $G_{\mc{C}}$, respectively. Subsequently, $V(G_{\mc{S}}) = X \cup Y$. The threshold of each $y_e \in Y, e \in E(G_{\mc{C}})$ is $3$, and the threshold value of each $x_u \in X, u \in V(G_{\mc{C}})$ is $k$. Lastly, we set $q = k + k(k - 1) / 2 = k(k+1) / 2$. This completes the construction of the \nminfpe{} instant. An example reduction is shown in Figure~\ref{fig:w1-hard}. To see that this is a parameterized reduction, we remark that the reduction can be carried out in $O(|V(G_{\mc{C}}) + E(G_{\mc{C}})|)$ time, which is fixed parameter tractable w.r.t. $k$. Furthermore, we have $q < k^3$.

\par We now claim that there is a clique of size $k$ in $G_{\mc{C}}$ if and only if $\mc{S}$ has a fixed point $C$ with Hamming weight $1 \leq H(C) \leq q$.

\begin{itemize}
    \item[($\Rightarrow$)] Let $V' \subseteq V(G_{\mc{C}})$ be the subset of vertices that form a $k$-clique in $G_{\mc{C}}$. We construct a fixed point $C$ of $\mc{S}$ by first setting vertices $x_u \in X, \forall u \in V'$ to state $1$. Furthermore, for each edge $e = (u, v) \in E(G_{\mc{C}})$ where $u, v \in V'$, we set the vertex $y_e \in Y$, which is the vertex that corresponds to $e$, to state $1$. All other vertices in $V(G_{\mc{S}})$ are set to state $0$ under $C$. Note that the number of state-$1$ vertices in $X$ are $Y$ is $k$ and $k(k-1) / 2$, respectively, thus, we have $H(C) = k + k(k-1)/2 = k(k+1) / 2 = q$.
    
    \par We now claim that $C$ is a fixed point. Let $C'$ be the successor of $C$. We first consider each vertex $x_u \in X$. If the state of $x_u$ is $1$ under $C$, then we have $u \in V'$ and by the construction above, $x_u$ has exactly $k-1$ neighbors in state $1$. Thus, $\tau_u$ is satisfied and $x_u$ remains state $1$ under $C'$. On the other hand, if $x_u$ is in state $0$, then $u$ is not in the $k$ clique of $G_{\mc{C}}$, and $x_u$ would have no neighbors in state $1$ under $C$. Thus, the state of $x_u$ remains $0$ in $C'$. A similar argument can be made for each vertex $y_e \in V(G_{\mc{S}})$. Overall, we conclude that $C$ is a fixed point.
    
    \item[($\Leftarrow$)] We first prove the following claim.
    
    \begin{statement}{Claim 3.2.1} \label{claim:6:10}
    Any non-zero fixed point $C$ of $\mc{S}$ has Hamming weight at least $q = k(k+1) / 2$.
    \end{statement}
    
    Let $x_u \in X$ be a state-1 vertex under $C$. Since the threshold of $x_u$ is $k$, at least $k-1$ of its neighbors are in state $1$. Let $y_e$ be a state-1 neighbor of $x_u$. Given that the threshold value of $y_e$ is $3$, both of its neighbors should also be in state $1$ under $C$. Therefore, under any non-zero fixed point, the number of $x_u \in X$ in state $1$ is at least $k$ and the number of $y_e \in Y$ in state 1 is at least $k(k-1) / 2$. Thus, the Hamming weight of $C$ is at least $k(k+1) / 2$. This concludes the claim.
    
    \par Now suppose there exists a fixed point $C$ with $1 \leq H(C) \leq q$, by the claim above, we must have $H(C) = q$. Note that the equality in the claim~3.2.1 holds only when the number of state-1 $x_u \in V(G_{\mc{S}})$ is exactly $k$ and the number of state-1 $y_e \in V(G_{\mc{S}})$ is exactly $k(k-1) / 2$, that is, the set of vertices $\{u \in V(G_{\mc{C}}), \text{$x_u$ is in state-1 under $C$}\}$ form a $k$ clique of $G_{\mc{C}}$.
\end{itemize}

Since \ts{clique} is \textbf{W[1]}-hard w.r.t. the natural parameter $k$, the theorem follows immediately.
\end{proof}

Theorem~(\ref{thm:w1}) implies that \nminfpe{} is not FPT w.r.t. to the natural parameter. Note that this does not exclude FPT results for other parameters, as we identify one such parameter in the next section.

\section{Approaches for Solving NMIN-FPE}
In this section, we consider several approaches for  tackling the hardness of \nminfpe{}. We start by identifying special cases where the problem can be solved efficiently.
We also present an integer linear programming (ILP) formulation that can be used to obtain optimal solutions for networks of reasonable sizes. Then we introduce a heuristic framework for \nminfpe{} that is useful in obtaining good (but not necessarily optimal) solutions in larger networks.

\subsection{Efficient algorithms for special classes}
\bigskip
\paragraph{Restricted classes.} We identify four special classes of problem instances where \nminfpe{} can be solved in polynomial time. Motivated by real-world scenarios, we consider the classic \textit{progressive threshold model}~\cite{kempe2003maximizing}, where once a vertex changes to state $1$, it retains the state $1$ for all subsequent time-steps. We also investigate special graph classes such as directed acyclic graphs and complete graphs. 

\begin{mybox2}
\begin{observation}\label{obs:monotone}
Let $\mc{S}$ be a $(\ts{Bool}, \ts{Thresh})$-SyDS. Given two configurations $C_1$ and $C_2$ of $\mc{S}$, let $C_1'$ and $C_2'$ be the successors of $C_1$ and $C_2$, respectively. If $C_1 \preceq C_2$, then $C_1' \preceq C_2'$.
\end{observation}
\end{mybox2}

\begin{mybox2}
\begin{theorem}{4.1} \label{thm:special_poly}
For (\ts{Bool, Thresh})-SyDSs, \nminfpe{} admits a polynomial time algorithm for any of the following restricted cases:

1. There exists at least one vertex with $\tau_v = 0$.

2. The SyDS uses the progressive threshold model.

3. The underlying graph is a directed acyclic graph.

4. The underlying graph is a complete graph.
\end{theorem}
\end{mybox2}

\begin{proof}
Let $\mc{S} = (\sydsG{}, \mc{F})$ be an arbitrary (\ts{Bool}, \ts{Thresh})-SyDS.

\begin{statement}{Claim 4.1.1}\label{claim1:nminfpe_p}
If there exists a vertex $v$ with $\tau_v = 0$, then \nminfpe{} can be solved in $O(m + n)$ time.
\end{statement}

 Let $V' = \{v \; : \; v \in V(G_{\mc{S}}), \tau_v = 0\}$ be the subset of vertices with thresholds $0$. Let $C_1$ be the configuration that consists of all $0$'s. We consider the evolution of $\mc{S}$ from $C_1$. Let $C_i$ denote the successor of the configuration $C_{i-1}$, $i \geq 2$, that is, $C_2$ is the successor of $C_1$, $C_3$ is the successor of $C_2$, etc. Following from the observation~\ref{obs:monotone}, we have $C_i \preceq C_{i+1}$, $i \geq 1$. Since $G_{\mc{S}}$ has a finite number of vertices, by recursion, $\mc{S}$ must reach a fixed point, denoted by $C_k$, from $C_1$ in $O(n)$ transitions. The evolution of $\mc{S}$ from $C_1$ to $C_k$ can be seen as a BFS process which takes $O(m + n)$ time.
 
 We now argue that $C_k$ is a nontrivial minimum fixed point of $\mc{S}$. It is easy to see that $C_k$ is non-zero since all vertices in $V'$ must be in state $1$ under $C_k$. Furthermore, for any non-zero fixed point $C$, we have $C_1 \preceq C$.  Since $C$ is a fixed point, the successor of $C$ is $C$ itself. Thus, by the observation~\ref{obs:monotone}, we have $C_i \preceq C$, $1 \leq i 
 \leq k$. It follows that all state-1 vertices in $C_k$ are also in state $1$ under $C$, and $H(C_k) \leq H(C)$. This concludes the claim 4.1.1.

For other cases below, we assume that $\tau_v > 0, \forall v \in V(G_{\mc{S}})$ without losing generality. 

\begin{statement}{Claim 4.1.2 }\label{claim2:nminfpe_p}
Under the progressive threshold model, \nminfpe{} can be solved in $O(mn + n^2)$ time.
\end{statement}

We label vertices in $V(\sydsG{})$ from $v_1$ to $v_n$ arbitrarily. The algorithm consists of $n$ iterations. At the $i$th iteration, $1 \leq i \leq n$, we construct an initial configuration $C_i$ by setting only vertex $v_i$ to state $1$, and all other vertices to state $0$. We then evolve the system from $C_i$ which always reaches to some fixed point $C_i^*$ in $O(n)$ time-step, since no vertices can flip from state $0$ back to state $1$. By repeating the above process for $n$ iterations, we get a collection of $n$ fixed points $\mc{I} = \{C^*_1, C^*_2, ..., C^*_n\}$ (they are not necessarily unique). Because of the progressive model, all $n$ fixed points are non-zero. We remark that a {\em{fixed point with minimum Hamming weight in $\mc{I}$ is a nontrivial minimum fixed point of the system}}, denoted by $C^*$. To see this, note that given any non-zero fixed point $C$ of $\mc{S}$, if the state of some $v_i \in V(G_{\mc{S}})$ is $1$ under $C$, then all state-$1$ vertices in $C^*_i$ must also be in state $1$ under $C$, thus, $H(C) \geq H(C_i^*) \geq H(C^*)$. 

\par As of running time, each iteration involves a full evolution of $\mc{S}$ from an initial configuration to a fixed point which can be seen as a BFS process taking $O(m + n)$ time. Thus, the overall running time is $O(mn + n^2)$. We remark that if $G_{\mathcal{S}}$ is connected, then the running time is $O(mn)$. This concludes the proof of the Claim~{4.1.2}.

\begin{statement}{Claim 4.1.3}\label{claim4:nminfpe_p}
\nminfpe{} can be solved in $O(mn + n^2)$ time for $\mc{S}$ when $G_{\mc{S}}$ is a directed acyclic graph.
\end{statement}

We first establish that if all vertices have thresholds greater than $1$, then $\mc{S}$ has \textbf{no} valid nontrivial minimum fixed point (In particular, the only fixed point of the system is the configuration that consists of all $0$'s). Given any fixed point $C$ of $\mc{S}$, let $v_1 \in V(G_{\mc{S}})$ be a vertex with $0$ in-degree. Note that such a vertex $v_1$ must exist, or else, $G_{\mc{S}}$ has at least one directed cycle. Since $\tau_{v_1} \geq 2 > d^-_{v_1} + 1$ where $d^-_v = 0$ is the in-degree of $v_1$, the state of $v_1$ is $0$ under $C$, and the input from $v_1$ to any one of its out-neighbors' local functions is $0$. Thus, we can only consider the subgraph induced on $V(G_{\mc{S}}) \setminus \{v_1\}$. Note that the resulting subgraph is also acyclic. Let $v_2 \in V(G_{\mc{S}}) \setminus \{v_1\}$ be the next vertex with $0$ in-degree. Similarly, $v_2$ must also have state $0$ under $C$. By recursion, we conclude that all vertices have state $0$ under $C$ if all vertices have thresholds greater than $1$. 

\par Suppose there exists at least one vertex with a threshold equals to $1$ (we have assumed that no vertices have threshold $0$) in $\mc{S}$. The algorithm in finding a nontrivial minimum fixed point is shown in Algorithm~\ref{alg:nminfpe_dag_supp}. We now prove its correctness. Let $C$ be any non-zero fixed point of $\mc{S}$. We claim that among all the vertices with state $1$ under $C$, one of them must have a threshold equal to $1$. For contradiction, suppose all the state-$1$ vertices have thresholds at least $2$. This implies that each state-$1$ vertex must have an in-neighbor who also has state $1$ under $C$. Since the graph is finite, the state-1 vertices form a cycle which contradicts $G_{\mc{S}}$ being acyclic. Thus, any non-zero fixed point $C$ of $\mc{S}$ must contain at least one state-$1$ vertex with a threshold equal to $1$. Let $v$ denote such a vertex, we then have $H(C) \leq H(C^*_v)$, where $C^*_v$ is the fixed point reach from the initial configuration with only $v$ being in state $1$. The correctness of the algorithm follows immediately. Lastly, a full evolution of $\mc{S}$ from an initial configuration to a fixed point can be treated as a BFS process which takes $O(m+n)$ time, thus, the running time of the algorithm is $O(mn + n^2)$.

\begin{statement}{Claim 4.1.4}
\nminfpe{} can be solved in $O(n^2 + m)$ time for $\mc{S}$ when $G_{\mc{S}}$ is a complete graph.
\end{statement}

Let $q$ be the number of distinct threshold values in $\mc{S}$. We partition the vertex set $V(G_{\mc{S}})$ into $q$ non-empty subsets $P = \{V_1, ..., V_p\}$, such that two vertices have the same threshold if and only if they are in the same subset $V_i$, $1 \leq i \leq q$. Moreover, subsets are in ascending order by thresholds, that is, $\tau_u < \tau_v$ if and only if $u \in V_i$, $v \in V_j$ and $1 \leq i < j \leq q$. Given that $G_{\mc{S}}$ is a complete graph, we make the following key observations. First, under a configuration $C$, the number of state-$1$ vertices in the closed neighborhood of each vertex is $H(C)$. Second, if $C$ is a fixed point and a vertex $u \in V_i$, $1 \leq i \leq q$ is in state $1$, then all vertices in the subsets $V_1, ..., V_i$ must be in state $1$.  

Under a configuration $C$, we call a state-$1$ vertex $v$ \textit{unsatisfied} if the number of state-$1$ vertices in $v$'s closed neighborhood is less than $\tau_v$, that is, $H(C) < \tau_v$. Overall, the algorithm starts with a configuration $I$ of all $0$'s and iteratively sets vertices to state $1$ until a there are no unsatisfied vertices. Specifically, at an iteration of the algorithm, let $V'$ be the subset of unsatisfied vertices under $I$, and let $\tau'$ denote the smallest threshold among all state-$0$ neighbors of vertices in $V'$. We set the states of all vertices with thresholds at most $\tau'$ to $1$ under $I$ and repeat the process until $V'$ is an empty set. The detailed pseudocode is given in Algorithm~\ref{alg:nminfpe_complete}. The while loop (line 5) in Algorithm~\ref{alg:nminfpe_complete} consists of at most $n$ iterations. The for loop from line 6 to 10 takes $O(n)$ time since we never set the same vertex to state $1$ twice. Furthermore, all other operations from line 11 to 15 take $O(n)$ time. Given that the system evolution at line $17$ takes $O(m + n)$ time, the running time of the Algorithm~\ref{alg:nminfpe_complete} is $O(n^2 + m)$.

We now claim the correctness of the algorithm. Consider the configuration $I$ (defined in the pseudocode) after the termination of the while loop in the algorithm~\ref{alg:nminfpe_complete}. It is easy to see that the thresholds of all state-$1$ vertices are satisfied under $I$. Follows from the monotonicity of threshold functions, $\mc{S}$ evolving from $I$ will always reach a non-zero fixed point $I^*$.

\par We argue that $I^*$ is minimum. Let $C$ be any non-zero fixed point, and let $\tau_{C} = \max\{\tau_v \; | \; \forall v \in V(\sydsG{}), C(v) = 1\}$ be the largest threshold over all state-$1$ vertices under $C$. Moreover, let $\tau_{I} = \max\{\tau_v \; | \; \forall v \in V(\sydsG{}), I(v) = 1\}$ be the largest threshold over all state-$1$ vertices under $I$. We must have $\tau_{I} \leq \tau_{C}$. For contradiction, suppose $\tau_{I} > \tau_{C}$. Let $V_j \in P$ be the subset in the partition where $\tau_v  = \tau_{C}, \forall v \in V_j$. Follows from $C$ being a fixed point, the while loop would have terminated when the states of all vertices in $V_j$ are set to one under $I$, implying that $\tau_{I} = \tau_{C}$ which yields a contradiction. It follows that $\tau_{I} \leq \tau_{C}$. Since $\sydsG{}$ is a complete graph, in any non-zero fixed point of $\mc{S}$, if $v \in V(\sydsG{})$ has state $1$, then all vertices $u$ with $\tau_u \leq \tau_v$ are in state $1$. Subsequently, all state-$1$ vertices in $I$ are also in state $1$ under $C$. It follows immediately that all state-$1$ vertices in $I^*$ are also in state $1$ under $C$, implying $H(I^*) \leq H(C)$. This concludes the claim 4.1.4. 
\end{proof}

\begin{algorithm}[H]
\SetKwInOut{Input}{Input}
\SetKwInOut{Output}{Output}
\SetAlgoLined
\LinesNumbered
\setstretch{0.5}
\Input{A (\ts{Bool}, \ts{Thresh})-SyDS 
$\mathcal{S}$ = $(G_{\mc{S}}, \mathcal{F})$ where $G_{\mc{S}}$ is a dag.}

\Output{A non-trivial minimum fixed point $C^*$}

 $V' = \{v : v \in V(G_{\mc{S}}), \tau_v = 1\}$ \\
 
 $min\_h = |V(G_{\mc{S}})| + 1$ \\
 
 \For{\textbf{each} $v \in V'$}
 {
    $C_v \gets$ the configuration where only $v$ is in state $1$ \\
    
    $C^*_v \gets$ the fixed point of $\mc{S}$ reached from $C_v$ \\
    
    \If{$H(C^*_v) < min\_h$}
    {
        $min\_h = H(C^*_v)$ \\
        
        $C^* = C^*_v$\\
    }
 }
 \Return{$C^*$}\\
 
 \caption{\texttt{\nminfpe{}\_DAG}($\mc{S}$)}
 \label{alg:nminfpe_dag_supp}
\end{algorithm}

\begin{algorithm}[H]
\SetKwInOut{Input}{Input}
\SetKwInOut{Output}{Output}
\SetAlgoLined
\LinesNumbered
\setstretch{0.5}
\Input{A (\ts{Bool}, \ts{Thresh})-SyDS $\mathcal{S} = (G_{\mc{S}}, \mathcal{F})$; $G_{\mc{S}}$ is a complete graph}

\Output{A non-trivial minimum fixed point $C^*$}
 
 $P \gets \{V_1, V_2, ..., V_p\}:$ a partition of $V(G_{\mc{S}})$ such that two vertices have the same threshold iff they are in the same subset, and $\tau_u < \tau_v$ iff $u \in V_i$, $v \in V_j$ and $1 \leq i < j \leq 1$. \\
 
 $V' \gets V_1$ \\
 
 $I \gets$ a configuration that consists of all zeros \\
 
 $j = k = 1$ \\
 
\While{$V' \neq \emptyset$}
{
    \For{$i$ \textbf{from} $j$ \textbf{to} $k$}
    {
        \For{$v \in V_i$}
        {
            $I(v) = 1$ \Comment{Change the state of $v$ to $1$ under $I$}
        }
     }

     $V' = \{v \in \bigcup_{l = 1}^k V_l \; : \; H(I) < \tau_v \}$
    
     $N' = \{u : \exists v \in V', (u, v) \in E(G_{\mc{S}}), u \notin \bigcup_{l = 1}^k V_l\}$ \\
    
     $\tau' = \min\{\tau_u \; | \; u \in N'\}$ \\
    
     $j = k + 1$\\
    
  $k =$ index of the subset $V_k \in P$ where $\tau_v = \tau', \forall v \in V_k$ \\
}

 $I^* \gets$ the fixed point of $\mc{S}$, evolved from $I$\\
 
 \Return{$I^*$}
 
 \caption{\texttt{\nminfpe{}\_complete\_graphs($\mc{S}$)}}
 \label{alg:nminfpe_complete}
\end{algorithm}

\bigskip
\paragraph{Fixed parameter tractability.} We further extend the solvability of \nminfpe{} and 
establish that \nminfpe{} is fixed parameter tractable w.r.t. the number of vertices with thresholds greater than $1$. Specifically, we develop a  quadratic kernelization algorithm that finds a nontrivial minimum fixed point in time  $O(2^k + n + m)$.

\begin{mybox2}
\begin{theorem}\label{thm:fpt} 
For (\ts{Bool},~\ts{Thresh})-SyDSs,
\nminfpe{} is fixed parameter tractable w.r.t. the parameter $k$ which is the number of vertices with thresholds greater than $1$.
\end{theorem}
\end{mybox2}

\begin{proof}
Given an instance $\mc{S} = (\sydsG{}, \mc{F})$, let $k$ be the number of vertices with thresholds greater than $1$. We present an algorithm for finding a nontrivial minimum fixed point of $\mc{S}$ that runs in fpt-time w.r.t. $k$.

\par Let $\mc{P} = \{CC_1, CC_2, ... CC_l\}$ be the collection of subsets of vertices with thresholds equal to $1$, such that each subset $CC_i \in \mc{P}$ forms a maximally \textit{connected components} that consists of only threshold-$1$ vertices; $l \geq 1$ is the number of such components. Observe that for each $CC_i, 1 \leq i \leq l$, if one vertex in $CC_i$ has state $1$ under some fixed point $C$, then all vertices in $CC_i$ are in state $1$ under $C$ (with the possibility that some other vertices outside of $CC_i$ are also in state $1$). Let $I_i$ be the configuration where $I_i(v) = 1$ if and only if $v \in CC_i$, that is, all vertices in $CC_i$ are in state $1$ and other vertices are in state $0$ under $I_i$. By the monotonicity of the threshold function, $\mc{S}$ evolving from $I_i$ must reach a fixed point, denoted by $C_i$, $1 \leq i \leq l$. Subsequently, we have a collection $\{C_1, C_2, ..., C_l\}$ of fixed points (not necessarily unique) where each $C_i$ corresponds a fixed point reached from the initial configuration $I_i$, $1 \leq i \leq l$. We remark that for any fixed point $C$, if there exists a vertex $v \in CC_i$ with state $1$ under $C$, then $H(C) \geq H(C_i)$ where equality holds only when $C = C_i$. 

\par Now consider a nontrivial minimum fixed point $C^*$ of $\mc{S}$. Let $V^* = \{v \in V(\sydsG{}) : C^*(v) = 1\}$ be the set of state-$1$ vertices under $C^*$. We distinguish between two cases, either (1) $\exists CC_i \in \mc{P}$, s.t. $V^* \cap CC_i \neq \emptyset$, or (2) $\forall CC_i \in \mc{P}$, s.t. $V^* \cap CC_i = \emptyset$.  If $\exists CC_i \in \mc{P}$, $V^* \cap CC_i \neq \emptyset$, by our previous remark, $C^* = C_i$.

\par Now suppose that $\forall CC_i \in \mc{P}$, $V^* \cap CC_i = \emptyset$, namely, all state-$1$ vertices under $C^*$ have thresholds greater than $1$. We can then find such a $C^*$ by removing vertices with thresholds equal to $1$ in $\sydsG{}$, leaving only the following subgraph: 
$$\sydsG{}' = \sydsG{}[V(\sydsG{}) \setminus \bigcup_{i=1}^l CC_i]$$
of $k$ vertices, where $\sydsG{}[V(\sydsG{}) \setminus \bigcup_{i=1}^l CC_i]$ is the subgraph induced on $V(\sydsG{}) \setminus \bigcup_{i=1}^l CC_i$. Since thresholds of vertices in $CC_i$ $1 \leq i \leq l$ are $1$, neighbors of vertices in $CC_i$ must have state $0$ under $C^*$. Therefore, to find an optimal solution $C^*$ within the subgraph $\sydsG{}'$, we need an additional constraint such that for each vertex $v \in V(\sydsG{}')$ that is adjacent to vertices in $CC_i$, $1 \leq i \leq l$, $v$ must have state $0$.

Overall, our algorithm in finding a nontrivial minimum fixed point is as follows. First, we construct the collection $\mc{P} = \{CC_1, CC_2,. ..., CC_i\}$ of maximally connected component that consist of only threshold-$1$ vertices. Then, we compute the collection $\mc{Q} = \{C_1, C_2, ... C_l\}$ of corresponding fixed points. Let $C' = \argmin\{H(C_i) : C_i \in \mc{Q}\}$ be a fixed point in $\mc{Q}$ of the minimum Hamming weight. Next, we construct the subgraph $\sydsG{}'$ and enumerate over all fixed points restricted to $\sydsG{}'$ to find the an optimal fixed point, denoted by $C''$. Lastly, the algorithm returns $\argmin \{ H(C'), H(C'') \}$. The correctness of the algorithm follows immediately from our arguments above.

As of the running time, observe that $\mc{P}$, $\mc{Q}$ and $C'$ can all be computed in $O(n + m)$ time. Moreover, since $|V(\sydsG{}')| = k$, $C''$ can be found in $O(2^k)$ time. Thus, the overall running time is then $O(2^k + n + m)$, ftp w.r.t. $k$. This concludes the proof.
\end{proof}

\subsection{Solving NMIN-FPE in general networks}
We make the following assumptions without losing generality. As shown in Theorem~\ref{thm:special_poly}, the problem \nminfpe{} can be solved efficiently when constant-1 vertices are presented in the system. To tackle the hardness of the problem, we thus assume that there are \textbf{no} constant-1 vertices in the system. Furthermore, we assume that the underlying graph is connected. Lastly, one can determine in polynomial time whether the only fixed point of a threshold-SyDS is the trivial fixed point (by starting the system with all vertices set to state 1, and the evolve the system). Therefore, the proposed heuristic does not explicitly consider this case and we assume the system has at least one non-trivial fixed point.

\paragraph{An ILP formulation.} Given a (\ts{Bool, Thresh})-SyDS~$\mc{S} = (\sydsG{}, \mc{F})$, our ILP  solves the \nminfpe{} problem by constructing a nonempty minimum-cardinality subset $V^* \subseteq V(\sydsG{})$ of vertices to set to state $1$, and all vertices not in $V^*$ to state $0$, such that the resulting configuration is a fixed point of $\mc{S}$. Let $x_v, \; v \in V(\sydsG{})$, be a binary variable where $x_v = 1$ if and only if vertex $v \in V^*$. Let $\Delta = max\_deg(G) + 2$ denote a constant that is greater than the maximum degree of $G$. Let $N(v), \; v \in V(\sydsG{})$ denote the closed neighborhood of $v$. Then an ILP formulation for \nminfpe{} is defined as follows in~(\ref{ilp_min}). An optimal solution to the ILP yields a set of state-$1$ vertices in a nontrivial minimum fixed point.
\begin{subequations}\label{ilp_min}
\begin{align}
& {\min}
& & \sum_{v \in V(\sydsG{})} x_v \label{ilp_min_c1_sup}\\
& \text{s.t.} & & \tau_v x_v  \leq \sum_{u \in N(v)} x_u, \;\;\; \forall v \in V(\sydsG{}) \label{ilp_min_c2_sup} \\
& & &  \sum_{v \in V(\sydsG{}) } x_v \geq 1 \label{ilp_min_c3_sup} \\
& & &  \Delta \cdot x_v + \tau_v \geq \sum_{u \in N(v)} x_u + 1, \; \; \; \forall v \in V(\sydsG{}) \label{ilp_min_c4_sup} \\
& & &  x_v \in \{0, 1\}, \;\;\; \forall v \in V(\sydsG{}) \label{ilp_min_c5_sup}
\end{align}
\end{subequations}

\begin{proof}
Under an optimal solution of the ILP, let $V^* = \{v \in V(\sydsG{}) : x_v = 1\}$ be the subset of vertices $v$ whose variable $x_v$ is set to $1$. Let $C^*$ be the configuration where $C^*(v)= 1$ if and only if $v \in V^*$. We now argue that $C^*$ is a fixed point. For a state-$1$ vertex $v$ under $C^*$, by the constraint~(\ref{ilp_min_c2_sup}), the number of state-$1$ vertices in the closed neighborhood of $v$ is at least $\tau_v$, thus, $v$ is fixed at state $1$. For a vertex $v$ with state $0$, by constraint~(\ref{ilp_min_c4_sup}), we have $\tau_v > \sum_{u \in N(v)} x_u$, that is, the number of state-$1$ vertices in the closed neighborhood is less than $\tau_v$ and the state of $v$ is fixed at $0$. Overall, we conclude that $C^*$ is a fixed point of $\mc{S}$. Since $V^*$ is minimum with $|V^*| \geq 1$, $C^*$ is a nontrivial minimum fixed point. 
\par By a similar argument, given a nontrivial fixed point $C^*$ of $\mc{S}$, the vertex set $V^* = \{v \in V(\sydsG{}) : C^*(v) = 1\}$ correspond to an optimal solution of the ILP that satisfies all the constraints. This concludes the proof.
\end{proof}

\paragraph{The greedy framework for heuristics.} As we have shown that \nminfpe{} is hard to approximate, one can only rely on heuristics to quickly find solutions for large networks. Given a (\ts{Bool, Thresh})-SyDS~$\mc{S} = (\sydsG{}, \mc{F})$, we propose a general framework that iteratively  constructs a fixed point by greedily setting vertices to state $1$. Specifically, the framework iterates over each vertex $u \in V(\sydsG{})$ and finds a fixed point seeded at $u$, which is denoted as $C_u$. In $C_u$, we have ($1$) $u$ is in state $1$, and ($2$) the Hamming weight of $C_u$ is heuristically the smallest. The algorithm then returns a fixed point $C_u$ with the smallest Hamming weight over all $u \in V(\sydsG{})$ as shown in Algorithm~\ref{alg:main}.

\begin{algorithm}[!htbp]
\SetKwInOut{Input}{Input}
\SetKwInOut{Output}{Output}
\SetAlgoLined
\setstretch{0.5}
\Input{SyDS $\mathcal{S} = (\sydsG{}, \mathcal{F})$}
\Output{A fixed point $C^* \in \{0, 1\}^n$}
 $best\_obj = |V(\sydsG{})|$ \\
 \For{$u \in V(G)$}
 {
    $C_u$ = \texttt{Greedy\_Seeded\_NM1FPE}$(\mc{S}, \; u)$ \\  
    \If{$H(C_u) < best\_obj$}
    {
        $best\_obj = H(C_u)$\\
        
        $C^* = C_u$\\
    }
 }
 \Return{$C^*$}\\
 
 \caption{\texttt{\textsc{NM1FPE}}$(\mc{S})$}
 \label{alg:main}
\end{algorithm}

We now introduce the greedy scheme to compute  $C_u, u \in V(\sydsG{})$. Let $A_u \subseteq V(\sydsG{})$ be a subset of state $1$ vertices under $C_u$. The scheme constructs $C_u$ by progressively adding vertices to $A_u$, which effectively set them to state $1$. Overall, the scheme has the following two key steps: (1) the construction of $A_u$ terminates when the \textit{fixed point condition} is met. Specifically, ($i$) for each vertex $v \in A_u$, the threshold $\tau_v$ is satisfied, that is, the number of $v$'s neighbors (including $v$ itself) in $A_u$ is at least $\tau_v$, and ($ii$) for each vertex $v \in V(\sydsG{}) \setminus A_u$, the number of $v$'s neighbors in $A_u$ is less than $\tau_v$; (2) vertices are added to $A_u$ based on a \textit{greedy selection method} specified by a heuristic. Under this framework, different heuristics can be obtained by
using different greedy selection strategies.

\paragraph{The fixed point condition.} We use a superscript to denote the iteration number. Initially, $A_u^{(0)}$ contains only $u$, and we actively add a new vertex to $A_u^{(k)}$ in each iteration $k \geq 1$. To determine if the fixed point condition is met at the $k$th iteration, we maintain a value $\delta^{(k)}$ that is the number of \textit{additional vertices} that need to be selected to satisfy the thresholds of vertices in $A_u^{(k)}$ at the $k$th iteration. Heuristically, $\delta^{(k)} = \sum_{v \in A_u^{(k)}} \Tilde{\tau}_v^{(k)}$ where $\Tilde{\tau}_v^{(k)} = \max\{0, \tau_v - |A_u^{(k)} \cap N(v)|\}$ is the \textit{residual threshold} of $v$ at the $k$th iteration. We can view $\Tilde{\tau}_v^{(k)}$ as the additional number of $v$'s neighbors that need to be selected to satisfy $\tau_v$. 

\par Given a vertex $w \in A_u^{(k)}$, we call $w$ \textit{unsatisfied} if $\Tilde{\tau}_w \neq 0$. Note that after adding a vertex $v$ to $A_u^{(k+1)}$, $v$ decreases the residual thresholds of its unsatisfied neighbors in $A_u^{(k)}$; yet, $v$ may remain unsatisfied.
Furthermore, we might ``passively'' satisfy the thresholds of some other vertices that are not in $A_u^{(k)}$. Let $A'$ denote such a set of ``passive'' vertices. We define $\epsilon_v^{(k)}$ to be the decrease of $\delta^{(k)}$ after selecting $v$ and $A'$. Thus, $\delta^{(k+1)}$ for the next iteration is computed by $\delta^{(k+1)} = \delta^{(k)} + \Tilde{\tau}_v - \epsilon_v^{(k)}$. Lastly, $A_u^{(k + 1)} = A_u^{(k)} \cup \{v\} \cup A'$. We have the following proposition.

\begin{mybox2}
\begin{proposition}
The fixed point condition is met at the $k$th iteration of the algorithm if and only if $\delta^{(k)} = 0$.
\end{proposition}
\end{mybox2}

The subroutine returns the fixed point $C_u$ where a vertex $v$ is in state $1$ iff $v \in A_u$. The pseudocode of the entire framework is presented under Algorithms~\ref{alg:main} and~\ref{alg:sub}.

\paragraph{Greedy selection strategies.} We now discuss a methodology for adding a new vertex $v$ to $A_u^{(k)}$. Observe that under any nontrivial minimum fixed point, the subgraph induced by the state-1 vertices is \emph{connected}. Thus, only the unselected neighbors of unsatisfied vertices, denoted by $B^{(k)}$, are candidate vertices at the $k$th iteration. Specifically, we greedily select a vertex $v = \argmin_{v \in B} \{obj(v)\}$ into $A_u^{(k)}$ where $obj(v)$ is an objective function specified by some heuristic. We now present objective functions for three heuristics. Let $\rho_v^{(k)}$ denote the number of unselected vertices that will be passively set to state 1 if $v$ is set to state $1$ at the $k$th iteration. The objective for the first heuristic \texttt{GreedyFull} is defined as $obj_1(v) = \Tilde{\tau}_v^{(k)} + \rho_v^{(k)} - \epsilon_v^{(k)}$ which considers $v$'s residual thresholds, the number of passive vertices, and the decrease of $\delta^{(k)}$. The objective functions of two other methods, namely \texttt{GreedyNP} and \texttt{GreedyThresh}, are simplifications of the first heuristic with $obj_2(v) = \Tilde{\tau}_v^{(k)} - \epsilon_v^{(k)}$ and $obj_3(v) = \Tilde{\tau}_v^{(k)}$, respectively. To speed up the execution time, we use \term{pruning} in the implementation of heuristics. Specifically, for each vertex $u$ enumerated in the heuristic framework, we keep track of the current optimal Hamming weight and actively terminate construction of $C_u$ if the accumulated Hamming weight of $C_u$ is larger than the current optimal value. In addition, we examine vertices in ascending order of their threshold values, thus heuristically attempting to find a small fixed point faster. To further simplify the \texttt{GreedyFull} algorithm, we propose \texttt{GreedySub}, which only examines the seeded fixed points for vertex $v$ if they have never been set to state one under $C_u$ for each vertex $u$ that is examined in previous iterations.

\begin{algorithm}[H]
\SetKwInOut{Input}{Input}
\SetKwInOut{Output}{Output}
\SetAlgoLined
\setstretch{0.5}
\Input{SyDS $\mathcal{S} = (\sydsG{}, \mathcal{F})$; vertex $u \in V(\sydsG{})$}
\Output{A fixed point $C_u$ seeded at vertex $u$}

$A' \gets $ The set of vertices being passively set to state $1$ if $s(u) = 1$\\

$A_u \gets \{u\} \cup A'$ \\

$B \gets \{N(u) \setminus A_u\}$  \Comment{Candidate vertices} 

$D = \{v \in A_u : \Tilde{\tau}_v \neq 0\}$  \Comment{Unsatisfied vertices}

$\delta = \max \{0, \tau_u - |A_u \cap N(u)|\}$ \\

 \While{$\delta \neq 0$}
 {
  $v^* = \argmin_{v \in B} \{obj(v)\}$ \Comment{Greedy selection}
  
  $\delta = \delta + \Tilde{\tau}_{v^*} - \epsilon_{v^*}$ \\
  
  $A' \gets $ The set of vertices being passively set to state $1$ after selecting $v^*$
  
  $A_u = A_u \cup \{v^*\} \cup A'$ \\
  
  Update sets $B$ and $D$ \\
}
    $C_u \gets$ the configuration with $v \in V(\sydsG{})$ in state $1$ iff $v \in A_u$\\
    
 \Return{$C_v$}\\
 
 \caption{\texttt{Greedy\_Seeded\_NM1FPE}$(\mc{S}, u)$}
 \label{alg:sub}
\end{algorithm}

\paragraph{The running time.} We first analyze the running time of Algorithm~\ref{alg:sub}. Remark that computing the set of vertices being passively put to state $1$ takes $O(m)$ time (line 1 and line 9) for each selected vertex. Further, the set of candidate vertices and unsatisfied vertices can be found in $O(n)$ time, corresponding to line 3, 4, and 11 of Algorithm~\ref{alg:sub}. If follows that the set of operations in line 1 to 5 and in line 8 to 11 take $O(m)$ time, given that $G_{\mc{S}}$ is connected and $n = O(m)$. Let $q$ denote the running time of a single greedy selection process at line $7$. The running time of one iteration of the while loop at line $6$ is then $O(m + q)$. Lastly, since at most $n$ vertices can be selected into $A_u$, the while loop at line $6$ consists of $O(n)$ iteration. It follows that the time complexity of Algorithm~\ref{alg:sub} is $O(nq + nm)$.

Algorithm~(\ref{alg:sub}), as a subroutine, is called exactly $n$ times by Algorithm~\ref{alg:main}.   Therefore, the time complexity of the proposed framework is $O(n^2 \cdot q + n^2m)$. Note that computing the objective values under the greedy rules of \texttt{GreedyNP/Thresh} takes constant time, since both $\Tilde{\tau}_v^{(k)}$ and $\epsilon_v^{(k)}$ are precomputed for each iteration $k \geq 1$. As for \texttt{GreedySub/Full}, observe that their objective value involves finding the set of passive vertices which takes $O(m)$ time. Subseqeuntly, the time complexity of \texttt{GreedyNP/Thresh} and \texttt{GreedySub/Full} are $O(n^2 m)$ and $O(n^3 m)$ respectively. We remark that via the pruning technique, the proposed algorithms empirically run much faster on networks of reasonable sizes, in contrast to their theoretical time complexity. 

\section{Experimental Results}
\par We conduct extensive experiments to investigate the performance of the heuristics under different experimental scenarios and stretch their performance boundaries. Overall, the results demonstrate high effectiveness of the heuristics in real-world networks.

\subsection{Experimental setup}
\textbf{Datasets.} We select the networks based on their sizes, diversity and application areas. Overall, we evaluate the heuristics on $13$ real-world networks from various domains 
(listed in Table~\ref{tab:networks}) 
and on Erd\"{o}s R\'{e}nyi  (Gnp) random networks.

\begin{center}
 \begin{tabular}{||l c c c c||} 
 \hline
 \textbf{Dataset} & \textbf{Type} & $n$ & $m$ & \textit{Max deg}\\ [0.5ex] 
 \hline 
  
  \texttt{router} & Infrastructure & 2,113 & 6,632 & 109 \\ \hline
  
  \texttt{power} & Infrastructure & 5,300 & 8,271 & 19\\ \hline
  
  \texttt{twitch} & Social & 7,126 & 35,324 & 720\\ \hline
  
  \texttt{retweet} & Social & 7,252 & 8,060 & 1,884\\ \hline
  
  \texttt{lastfm} & Social & 7,624 & 27,806 & 216 \\ \hline
  
  \texttt{arena} & Social & 10,680 & 24,316 & 205 \\ \hline
  
  \texttt{gnutella} & Peer-to-Peer & 10,876 & 39,994 & 103\\ \hline
  
  \texttt{auto} & Infrastructure &11,370 & 22,002 & 2,312\\ \hline
  
  \texttt{astroph} & Coauthor &17,903 & 196,972 & 504  \\ \hline
  
  \texttt{condmat} & Coauthor & 21,363 & 91,286 & 279\\ \hline
   
  \texttt{facebook} & Social & 22,470 & 170,823 & 709\\ \hline
  
  \texttt{google+} & Social & 23,613 & 39,182 & 2,761 \\ \hline
  
  \texttt{Deezer} & Social & 28,281 & 92,752 & 172 \\ \hline
\end{tabular}

\captionof{table}{List of networks}
\label{tab:networks}
\end{center}

\paragraph{Heuristics and baselines.} We evaluate the performance of the proposed greedy heuristics by comparing with the following baselines:  
(1) \texttt{DegDis}: a minimization version of the selection method proposed in ~\cite{chen2009efficient};
(2) \texttt{Random}: select vertices randomly. We also consider other methods that select vertices with a smallest value of the metrics: (3) \texttt{Pagerank}, (4) \texttt{Distance} (closeness centrality) which are also widely used by others as baselines~\cite{yao2015topic, kempe2003maximizing}. 

\paragraph{Experimental scenarios.} We consider the following three cases in investigating the effectiveness of the above natural heuristics. (1) \textit{Random thresholds}, (2) \textit{Uniform thresholds}, and (3) \textit{Gnp networks with increasing sizes}. The details of each setting are given in later sections.

\paragraph{Evaluation metric.} We use the \textit{approximation ratio} $\gamma = obj / OPT$ as the evaluation metric, where $OPT$ is the optimal objective (the Hamming weight of the fixed point) of a problem instance computed by solving the proposed ILP using \texttt{Gurobi}, and $obj$ is the objective value returned by a heuristics. We remark that $\gamma \geq 1$ and \textit{an algorithm with a lower $\gamma$ gives a better solution}.

\paragraph{Machine and reproducibility.} All experiments were performed on Intel Xeon(R) Linux machines with 64GB of RAM. Our source code (in \texttt{C++} and \texttt{Python}), documentation, and datasets are provided as supplementary materials.

\subsection{Experimental results}
In this section, we present the results of the heuristics under three experimental scenarios.

\paragraph{Random thresholds.} We first study the scenario where threshold values of vertices are assigned randomly in the range $[3, deg(v) + 1]$. This construction guarantees that there are no \textit{constant-$1$} vertices, as the \nminfpe{} would become efficiently solvable in that case by Theorem~\ref{thm:special_poly}. The random threshold assignment is a way to cope with the incomplete knowledge of the actual threshold values of vertices~\cite{kempe2003maximizing}. 

\par
\begin{figure*}[!h]
  \centering
    \includegraphics[width=1\textwidth]{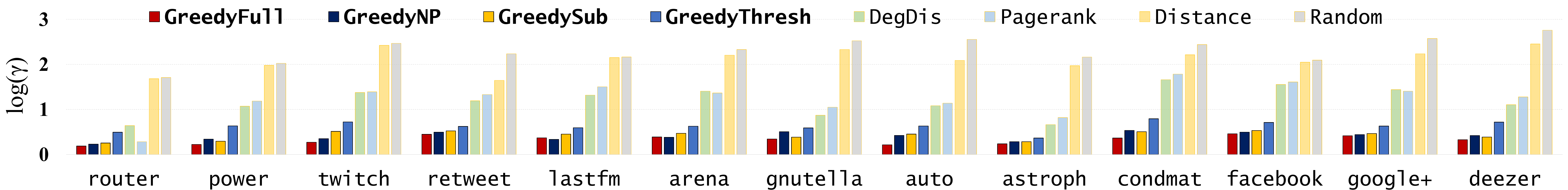}
     \caption{\textbf{The approximation ratios (lower is better) under random threshold setting.} The $y$ axis denotes the approximation ratios of heuristics under the $\log_{10}$ scale. The results are averaged over $10$ initializations of threshold assignments.}
\label{fig:rand}
\end{figure*}

 The results are averaged over $10$ initializations of threshold assignments, shown in Fig.~\ref{fig:rand} \textbf{under the $\log_{10}$ scale}. Overall, we observe that the proposed \texttt{Greedy} family significantly outperforms other baselines. Specifically, the averaged approximation ratios (over all networks) of \texttt{GreedyFull/NP/Sub} are less than $3$,
which is over $10$ times better than those of baselines on most networks. Within the \texttt{Greedy} family, the simplest heuristic \texttt{GreedyThresh} shows the lowest performance, with an averaged approximation ratio of $4.34$. Nevertheless, we remark that this empirically constant ratio is significantly better than that of other baselines. Further, \texttt{GreedyThresh}, 
 which is the most efficient among all its counterparts,
finds solutions in minutes.

\paragraph{Uniform thresholds.}
We assign all vertices the same threshold $\tau > 3$. We consider different values of the uniform threshold $\tau$ to study how the heuristics perform. Note that as $\tau$ increases, more constant-$0$ vertices emerge in the system because their threshold values are larger than their degrees. Thus, the uniform-threshold setting pushes the performance limits of algorithms by ($1$) setting the thresholds of vertices to be indistinguishable and ($2$) introducing a considerable number of constant-$0$ vertices which makes it harder for the heuristics to even find a feasible solution.

\begin{figure*}[!h]
  \centering
    \includegraphics[width=1\textwidth]{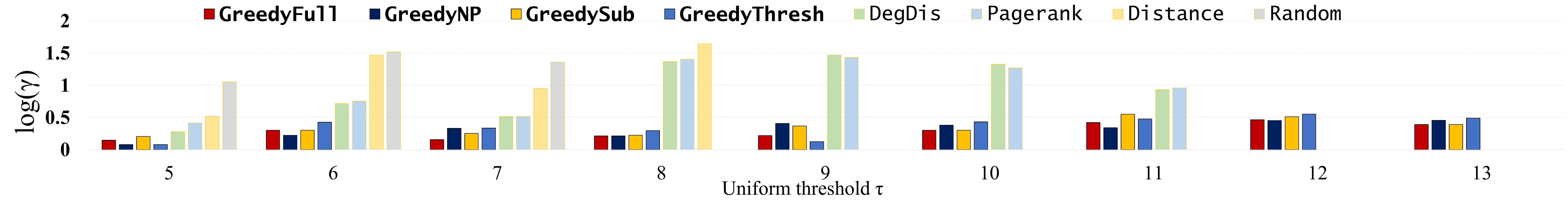}
    
     \caption{\textbf{The approximation ratios (lower is better) of heuristics on \texttt{google+} network under uniform threshold setting.} The $y$ axis shows the approximation ratios of heuristics under the $\log_{10}$ scale. Results of several heuristics (\texttt{DegDis}, \texttt{Pagerank}, \texttt{Distance}, \texttt{Random}) are absent for some instances because they failed to find valid fixed points.}
\label{fig:google}
\end{figure*}

We first present analyses on instances of the same network with different uniform thresholds~$\tau$. Due to page limits, we show results for the \texttt{Google+} network in Fig.~\ref{fig:google}; the results are similar 
for all other networks where the \texttt{Greedy} family outperforms the other heuristics in terms of approximation ratios. In particular, the averaged ratios for heuristics in the \texttt{Greedy} family are all less than $3$ for \texttt{Google+} network, with \texttt{GreedyThresh} having the highest averaged ratio (lower is better) of $2.41$. We also observe when the uniform threshold $\tau$ is large enough, many natural heuristics failed to even find a valid solution due to the presence of a large number of constant-$0$ vertices. This experiment demonstrates the high effectiveness of the proposed framework when vertices have indistinguishable thresholds.

\par
\begin{figure*}[!h]
  \centering
    \includegraphics[width=1\textwidth]{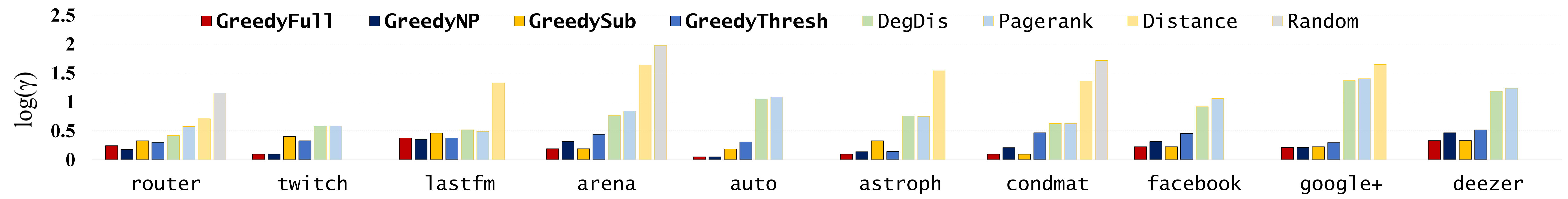}
    
     \caption{\textbf{The approximation ratios (lower is better) of heuristics on networks under uniform threshold setting where $\tau = 8$.} The $y$ axis shows the approximation ratios of heuristics under the $\log_{10}$ scale. Results of several heuristics (\texttt{DegDis}, \texttt{Pagerank}, \texttt{Distance}, \texttt{Random}) are absent for some instances because they failed to find valid fixed points.}
     
\label{fig:uni-8}
\end{figure*}

 Next, we fix the uniform threshold $\tau$ and analyze the heuristics across all networks. We have investigated different values of $\tau$ from $3$ to $20$, where the \texttt{Greedy} family again outperforms the other heuristics on most instances, producing results that are over 10 times better than baselines. Due to the page limit, we show the results for $\tau = 8$ in Fig.~\ref{fig:uni-8}. Note we omit the results for networks \texttt{power} and \texttt{peer} because they have no feasible solutions when $\tau = 8$.

\paragraph{Gnp networks.} We study the heuristics on Erd\"{o}s R\'{e}nyi networks with sizes up to $1,000$ where thresholds are assigned randomly. We observe that the ratios of \texttt{GreedyFull/NP/Thresh} are all over $40$, and the ratios of \texttt{GreedySub} are as high as $148.48$. We remark that in a gnp network, the state-$1$ vertices in a nontrivial minimum fixed point are often surrounded by vertices with similar thresholds and similar degrees. Our results suggest a limitation of the proposed framework, that is, when networks exhibit uniformity at the vertex level, the heuristics might not correctly choose state-$1$ vertices to construct a minimum fixed point. Note that such results are expected since \nminfpe{} is hard to approximate. We remark that the \texttt{Greedy} family still outperforms baselines on Gnp networks. 

\paragraph{Efficiency.} Our results demonstrate that the proposed \texttt{Greedy NP/Sub/Thresh} are more efficient than the ILP solver \texttt{Gurobi} for the tested scenarios. In Table~\ref{tab:time}, we show the runtime of 
\texttt{Greedy NP/Sub/Thresh} and \texttt{Gurobi} solver on the two largest networks under the random threshold scenario.

The heuristics achieve high efficiency by the pruning technique. For some networks, \texttt{Gurobi} runs comparably fast, usually within $5$ minutes. Nevertheless, \texttt{Gurobi} uses parallelization mechanisms such as multithreading (over $30$ threads are used on each instance) whereas our heuristics achieve high efficiency while executing in serial mode. 

\begin{center}
 \begin{tabular}{||l c c c c||} 
 \hline
 \textbf{Network} & \texttt{GreedyNP} & \texttt{GreedySub} & \texttt{GreedyThresh} & \texttt{Gurobi}\\ [0.5ex] 
 \hline 
 \texttt{Google+} & \textbf{26.13}s & 263.64s & 32.16s & 1153.04s \\ \hline
 \texttt{Deezer} & 104.76s & 395.89s & \textbf{96.22}s & 794.89s \\ \hline
\end{tabular}

\captionof{table}{Execution times of heuristics in seconds.}
\label{tab:time}
\end{center}

\section{Conclusions and Future Work}
In this paper, we study the  \nminfpe{} problem from both the theoretical and empirical points of view. We establish the computational hardness of the problem and propose effective algorithms for solving \nminfpe{} under special cases and general graphs. Our results point to a new way of quantifying system resilience against the diffusion of negative contagions and a new approach to tackle the influence minimization problem. One limitation of the proposed heuristics  is the cubic time complexity. Thus, a future direction is to develop more efficient methods for solving \nminfpe{}. Another promising direction is to approximate \nminfpe{} under restricted graph classes such as regular graphs. Lastly, we want to extend the model to multi-layer networks and investigate problems in this new domain.

\bibliographystyle{plain}
\bibliography{bib}

\end{document}